\documentclass[twocolumn]{aastex631}
\usepackage{newtxtext,newtxmath}

\usepackage{booktabs}
\usepackage[T1]{fontenc}

\DeclareRobustCommand{\VAN}[3]{#2}
\let\VANthebibliography\thebibliography
\def\thebibliography{\DeclareRobustCommand{\VAN}[3]{##3}\VANthebibliography}

\usepackage{graphicx}	
\usepackage{amsmath}	
\usepackage{amssymb}	
\usepackage{color}
\usepackage{xcolor}

\begin{document}

\title{Metastability in Emergent Dark Energy: A New Framework Confronting Cosmological Observations}

\author{Xiaolei Li}

\affiliation{College of Physics, Hebei Normal University, Shijiazhuang 050024, China}
\author{Tonghua Liu}

\affiliation{School of Physics and Optoelectronic, Yangtze University, Jingzhou 434023, China}

\author{Tian-Nuo Li}
\affiliation{Liaoning Key Laboratory of Cosmology and Astrophysics, College of Sciences, Northeastern University, Shenyang 110819, China}

\author{Guo-Hong Du}
\affiliation{Liaoning Key Laboratory of Cosmology and Astrophysics, College of Sciences, Northeastern University, Shenyang 110819, China}

\author{Arman Shafieloo}
\affiliation{Korea Astronomy and Space Science Institute, Daejeon 34055, Korea}
\affiliation{University of Science and Technology, Yuseong-gu 217 Gajeong-ro, Daejeon 34113, Korea}

\author{Marek Biesiada}
\affiliation{National Centre for Nuclear Research, Pasteura 7, 02-093
Warsaw, Poland}

\email{liutongh@yangtzeu.edu.cn}
\email{shafieloo@kasi.re.kr}
\email{Marek.Biesiada@ncbj.gov.pl}

\begin{abstract}
We propose the Metastable Emergent Dark Energy (MEDE) model, a novel phenomenological extension of the Phenomenological (PEDE) and Generalized (GEDE) Emergent Dark Energy frameworks, in which dark energy exhibits a transitionary behavior, appearing at late times and vanishing toward the future. This model naturally enables a smooth crossing of the phantom divide line in the dark energy equation of state, as hinted at by recent observations. The MEDE model is defined by a hyperbolic tangent dark energy equation of state $w(z)=-1-\Delta\tanh[\log_{10}((1+z)/(1+z_t))]$, introducing only two free parameters, the transition redshift $z_t$ and the variation amplitude $\Delta$, allowing both the emergent and transitionary behavior of dark energy. We constrain the MEDE model using a combined dataset of Planck CMB, DESI DR2 BAO, and different compilations of Type Ia supernovae, obtaining $z_t=0.425^{+0.084}_{-0.120}$ and $\Delta =0.87^{+0.29}_{-0.35}$ (for CMB+DESI+PantheonPlus), indicating a statistically significant deviation from the cosmological constant. Statistical comparisons show that the MEDE model is preferred over $\Lambda$CDM by the combined dataset, with $\Delta \rm DIC_{ MEDE-\Lambda CDM}= -9.29$. The MEDE model performs comparably to the CPL dynamical dark energy parametrization ($\Delta \rm DIC_{MEDE-CPL} = 0.74$), with no strong statistical distinction from CPL using current data. Notably, MEDE preserves the success of $\Lambda$CDM in describing early-universe physics and naturally accommodates the phantom-crossing signature indicated by the latest low-redshift observations. The MEDE scenario provides a compelling dark energy phenomenology that may guide us toward interesting theoretical implications. 
\end{abstract}




\section{Introduction}\label{sec:intro}
The standard $\Lambda$CDM cosmological model provides an excellent fit to a wide range of astrophysical observations, establishing the cornerstone of modern precision cosmology. Yet, it faces several persistent challenges that motivate the exploration of alternative dark energy paradigms. In particular, the Hubble tension, a $4$–$5\sigma$ discrepancy between early- and late-universe measurements of the Hubble constant and the $\sigma_8$ tension concerning the amplitude of matter clustering indicate potential limitations in the standard model's description of dark energy~\citep{Verde:2019ivm,DiValentino:2020zio,DiValentino:2022fjm,DiValentino:2021izs,Guo:2018ans,Hu:2023jqc,Kamionkowski:2022pkx,Vagnozzi:2019ezj,Vagnozzi:2023nrq,Abdalla:2022yfr}.

Recent results from the Dark Energy Spectroscopic Instrument (DESI) DR2~\citep{DESI:2025zgx} have strengthened the case for dynamical dark energy, reporting a $2.5\sigma$ indication that the dark energy equation of state may cross the phantom divide $w = -1$. This hint, together with earlier indications from other cosmological probes, has motivated a surge of theoretical activity exploring dynamical dark energy scenarios beyond the simple cosmological constant of $\Lambda$CDM~\citep{Giare:2024gpk,Giare:2024smz,Giare:2024oil,Li:2025cxn,Li:2024qus,Li:2025ula,Li:2024qso,Jiang:2024viw,Yang:2025uyv,Pan:2025qwy,Braglia:2025gdo,Li:2025owk,2025arXiv250321652S,Liu:2025mub,Liu:2025myr,Li:2026xaz,Li:2025ops,Fazzari:2025lzd,Ozulker:2025ehg,Wolf:2025jed,Wolf:2025acj,Du:2025xes,Du:2026cly,Teixeira:2025czm,vanderWesthuizen:2025rip,Cheng:2025lod,Silva:2025hxw,Pedrotti:2025ccw,Wang:2026kbg,Zhou:2025nkb,Zhang:2025dwu,RoyChoudhury:2025iis,Luciano:2025elo,Paliathanasis:2026ymi,Cheng:2025yue,DESI:2025fii,Keeley:2025stf,DESI:2025wyn}. {However, as cautioned in~\citet{Keeley:2025rlg}, the evidence for phantom crossing is not yet definitive. Moreover, even if current data continue to favor such behavior, realizing a crossing of $w = -1$ poses a theoretical challenge, as it cannot be achieved within simple quintessence or phantom models alone.}

The emergent dark energy paradigm offers a compelling alternative by suggesting that dark energy remained subdominant during most of the cosmic history, only emerging as a dominant component in the late universe. This idea was first realized phenomenologically in the Phenomenological Emergent Dark Energy (PEDE) model, which introduced a smooth transition in dark energy density while preserving the parameter count of $\Lambda$CDM \citep{2019ApJ...883L...3L}. To allow for more 
flexibility in the transition shape, the Generalized Emergent Dark Energy (GEDE) model was subsequently proposed, incorporating an additional parameter controlling the steepness of emergence \citep{2020ApJ...902...58L,2024MNRAS.533.1865L}.  {Recent studies have revisited the PEDE model and its generalized version (GEDE) using the latest cosmological data, finding that while PEDE is disfavored by combined datasets \citep{Hernandez-Almada:2024ost}, GEDE remains consistent with observations and is mildly favored depending on the supernova calibration \citep{Sharma:2025qmv}; moreover, the preference for dynamical dark energy over a cosmological constant may be sensitive to the choice of priors \citep{Payeur:2024dnq}.} However, both PEDE and GEDE share a fundamental limitation: they do not allow the dark energy equation of state to cross the phantom divide at $w = -1$, restricting their ability to describe transitions between quintessence ($w > -1$) and phantom ($w < -1$) regimes.

This empirical motivation leads us to propose the Metastable Emergent Dark Energy (MEDE) model, which extends the PEDE and GEDE frameworks by allowing smooth phantom crossing while maintaining theoretical consistency. The MEDE parametrization emerges naturally from scalar-field dynamics with non-canonical kinetic terms, particularly in quintom scenarios where phantom and quintessence behaviors coexist. 
Crucially, MEDE preserves the attractive features of its predecessors: dark energy remains subdominant at early times (safeguarding the well-established high-redshift cosmology), and the model introduces only two additional parameters beyond $\Lambda$CDM--the transition redshift $z_t$ and the amplitude of variation $\Delta$--matching the complexity of widely used dynamical dark energy parametrizations like Chevallier--Polarski--Linder (CPL)\citep{Chevallier:2000qy,Linder:2002et}.  {The CPL parametrization describes the dark energy equation of state as $w(z) = w_0 + w_a {z}/{(1+z})$, where $w_0$ is the present-day value and $w_a$ characterizes its time evolution. While CPL is a widely used empirical tool for describing dark energy evolution, it remains agnostic about the underlying physics. In contrast, the emergent dark energy paradigm offers a specific conceptual framework in which dark energy emerges only at late times. MEDE provides a phenomenological realization of this picture that, unlike PEDE and GEDE~\citep{2019ApJ...883L...3L,2020ApJ...902...58L}, allows for phantom-crossing. }

In this work, we perform a comprehensive test of the MEDE framework by combining the latest Planck CMB data~\citep{Aghanim:2018eyx}, DESI BAO measurements~\citep{DESI:2025zgx}, and the PantheonPlus Type Ia supernovae sample. Our analysis investigates the model's ability to simultaneously address cosmological tensions while accounting for the possible $w$-crossing behavior suggested by DESI.

The paper is organized as follows: Section~\ref{sec:model} presents the theoretical framework of the MEDE model; Section~\ref{sec:data_and_methodology} describes the observational data and statistical methods used in our analysis; Section~\ref{sec:results} details the observational constraints and compares them with standard $\Lambda$CDM and CPL parameterization; and Section~\ref{sec:con} discusses the implications of our findings for fundamental cosmology and future observational programs.

\section{Metastable Emergent Dark Energy Model} \label{sec:model}

In a spatially flat Friedmann-Lemaitre-Robertson-Walker (FLRW) universe, the expansion history is governed by the Hubble parameter:
\begin{equation}
H^2(z) = \frac{8\pi G}{3} \left( \rho_{m0}(1+z)^3 + \rho_{\text{de}}(z) + \rho_{r0}(1+z)^4 \right),
\end{equation}
where $\rho_{m0}$, $\rho_{r0}$, and $\rho_{\text{de,0}}$ are the present-day energy densities of matter, radiation, and dark energy, respectively. The evolution of dark energy density for MEDE model can be written as:
\begin{equation} \label{eq:rhode}
\rho_{\text{de}}(z) = \rho_{\text{de,0}} \left[ \frac{\cosh\left( \log_{10} \left( \frac{1+z}{1+z_t} \right) \right) }{\cosh \left( \log_{10}\left( \frac{1}{1+z_t} \right) \right) } \right]^{-3\Delta \ln10},
\end{equation}
which directly leads to a dynamical equation of state:
\begin{equation} \label{eq:wz}
w(z) = -1 - \Delta \tanh \left( \log_{10}\left( \frac{1+z}{1+z_t} \right) \right).
\end{equation}
This formulation ensures that all equations reduce to the $\Lambda$CDM limit when $\Delta = 0$, preserving the  well-tested success of the standard model at the background level.

To consistently compute cosmological observables like the CMB and matter power spectra, the background dynamics must be complemented by a description of linear perturbations. We consider the linear perturbations of MEDE within the comoving Newtonian gauge. The continuity and Euler equations, governing the evolution of dark energy inhomogeneities, are derived from the first-order perturbations of the conserved stress-energy-momentum tensor:
\begin{align}\label{eq:rhode}
 \delta^{\prime}_{\rm de}=& -(1+w_{\rm de}) \left(\theta_{\rm de} - 3 \phi^{\prime} \right)-3 \mathcal{H} \delta_{\rm de} (c^2_{\rm s,de} - w_{\rm de}) 
 \nonumber\\
 &- 9 (1+w_{\rm de})(c^2_{\rm s,de} - c^2_{\rm a,de})\mathcal{H}^2 \frac{\theta_{\rm de}}{k^2},
\end{align}
\begin{equation}\label{eq:thetade}
\theta^{\prime}_{\rm de}=-(1-3 c^2_{\rm s,de}) \mathcal{H} \theta_{\rm de}  + \frac{c^2_{\rm s,de}}{1+w_{\rm de}}k^2 \delta_{\rm de} + k^2\psi.
\end{equation}
Here, primes denote differentiation with respect to conformal time, $\mathcal{H}\equiv a^{\prime}/a$ is the conformal Hubble parameter, $\psi$ and $\phi$ are the Bardeen metric potentials, and $k$ is the wavevector magnitude. The density and velocity perturbations for MEDE are $\delta_{\rm de}$ and $\theta_{\rm de}$, respectively, while $c_{\rm s,de}$ and $c_{\rm a,de}$ denote the sound speed and adiabatic sound speed in dark energy fluid. We fix $c^2_{\mathrm{s,de}} = 1$, which is a common choice that prevents unphysical instabilities on small scales and influences the growth of structure, which is crucial for addressing the $\sigma_8$ tension. The term $w_\mathrm{de}$ in these equations is given by Eq.~\eqref{eq:wz}, intrinsically linking the perturbation evolution to the distinctive background dynamics of the MEDE model.

The MEDE model is thus motivated by a coherent set of theoretical and phenomenological considerations:

\begin{enumerate}
\item  {Dynamical Dark Energy and the Phantom Divide}\
The model naturally incorporates a crossing of the phantom divide ($w = -1$), a feature suggested by observations like DESI but challenging for simple quintessence or phantom models. This is inherently encoded in the $\tanh$ form of $w(z)$, with the transition centered at $z_t$.
\item  {Emergent and Structured Evolution}\\
The ``emergent'' nature refers to dark energy becoming dominant only at late times. However, unlike a cosmological constant, its density evolves structurally: for $\Delta > 0$, it transitions from phantom-like ($w < -1$) at $z > z_t$ to quintessence-like ($w > -1$) at $z < z_t$, gracefully modifying the expansion history.

\end{enumerate}

In summary, the MEDE model represents a minimal, physically motivated extension to $\Lambda$CDM. It combines  self-consistently a specific background evolution with its perturbation theory, providing a testable framework for resolving key tensions in modern cosmology. Future data from DESI, Euclid, and CMB-S4 will critically test its viability.

\begin{figure*}
\centering
\includegraphics[width=1\textwidth]{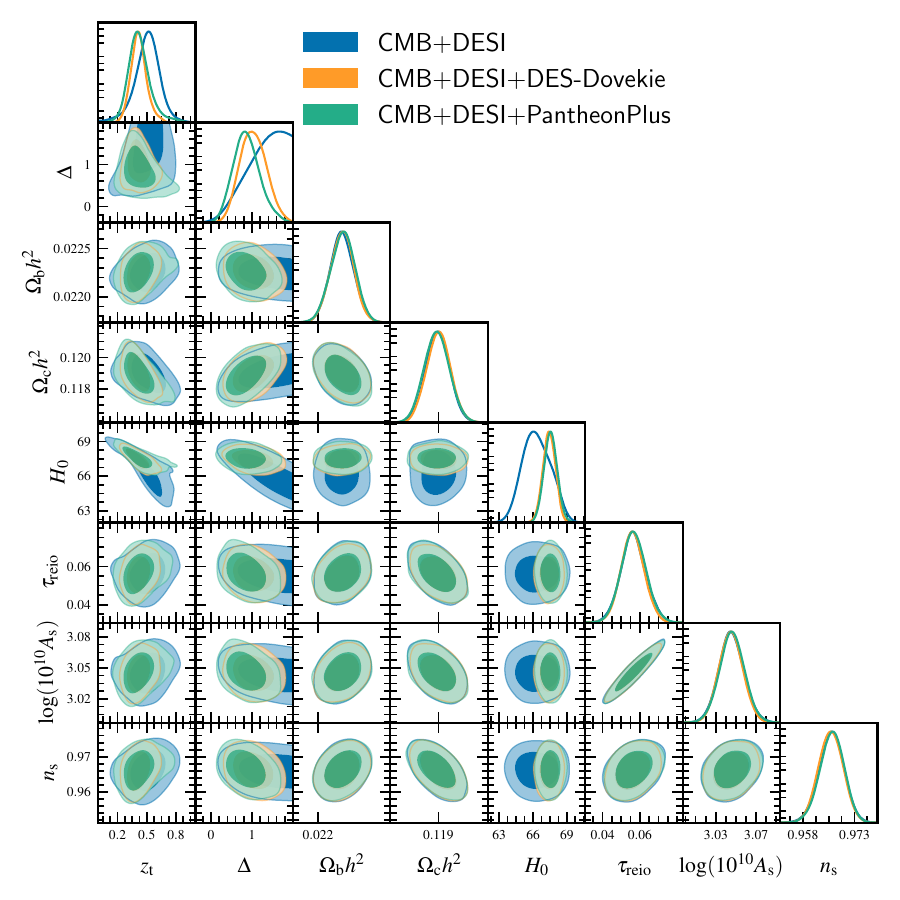}
\caption{Constraints on the cosmological parameters of the MEDE model.
The panels show the one-dimensional posterior distributions and two-dimensional joint confidence regions (68\% and 95\%) for the core cosmological and MEDE model parameters, derived from the combination of CMB, DESI BAO, DES-Dovekie, and PantheonPlus supernova data.}
\label{Fig:mede1}
\end{figure*}

\section{Observational data }\label{sec:data_and_methodology}
The datasets used in the analyses are described below:
We present a comprehensive cosmological analysis combining multiple recent observational datasets to constrain the MEDE model. Our dataset compilation includes:

 {BAO}: We utilize the latest BAO measurements from the DESI DR2 \citep{2025arXiv250314738D}, which represents the most precise BAO compilation to date. The dataset combines four distinct tracer populations: luminous red galaxies, emission line galaxies, quasars, and the Lyman-$\alpha$ forest. These measurements provide constraints on three fundamental distance ratios: the transverse comoving distance $D_M/r_d$, the Hubble distance $D_H/r_d$, and the volume-averaged distance $D_V/r_d$, where $r_d$ denotes the sound horizon at the drag epoch. 
Our analysis properly accounts for the complete covariance structure, including the significant cross-correlation coefficient $r_{M,H}$ between transverse and radial distance measurements, which typically ranges from 0.3 to 0.5 across different redshift bins. This comprehensive BAO dataset is referred to as  {DESI} throughout our work.

 {SN Ia}: We employ two independent SN Ia compilations to constrain the late-time expansion history:
i). We use the full Pantheon Plus sample \citep{2022ApJ...938..113S}, consisting of 1,701 high-quality light curves from 1,550 spectroscopically confirmed SN Ia spanning $0.01 < z < 2.26$. To mitigate potential systematics from low-redshift calibration uncertainties, we exclude the calibration subsample at $z < 0.01$ and rigorously incorporate the full covariance matrix$\footnote{\url{https://github.com/PantheonPlusSH0ES/DataRelease}}$, which captures both statistical errors and systematic correlations. The final sample of 1,485 SN Ia provides a robust Hubble diagram for cosmological inference;
ii). We employ the fully recalibrated DES-Dovekie supernova sample \citep{DES:2025sig}, derived from a comprehensive re-analysis of the DES 5-year SN Ia. This sample incorporates several key improvements from \cite{2024ApJ...973L..14D}, including an enhanced photometric cross-calibration, the use of recent white dwarf observations for calibration across surveys, a retrained SALT3 light-curve model, and a corrected host galaxy colour law. The final DES-Dovekie dataset consists of approximately 1600 likely Type Ia supernovae from DES and about 200 low-redshift supernovae from external surveys, providing a robust foundation for constraining cosmological models.
These datasets are labeled  {PantheonPlus} and  {DES-Dovekie}, respectively.

 {CMB}: Our CMB constraints are derived from the \textit{Planck} 2018 legacy data \citep{Aghanim:2018eyx}, including high-$\ell$ temperature and polarization spectra (TT+TE+EE) and Low-$\ell$ temperature (TT) and polarization (EE) likelihoods.
These measurements precisely determine the acoustic peak structure and are sensitive to fundamental cosmological parameters including the baryon density, matter content, and expansion history.
Additionally, we include CMB lensing constraints from the combined \textit{Planck} PR4 (NPIPE) reconstruction \citep{PlanckPR4} and Atacama Cosmology Telescope DR6 results \citep{ACTDR6lensing1,ACTDR6lensing2,ACTDR6lensing3}. The lensing potential power spectrum $C_\ell^{\phi\phi}$ probes the integrated matter distribution and provides complementary constraints on late-time structure growth.
The combined dataset, denoted as  {CMB}, comprises \textit{Planck} 2018 TT, TE, EE spectra together with the \textit{Planck}+ACT lensing likelihood.

The theoretical framework of the MEDE model, as defined by its background evolution (Eqs.~\ref{eq:rhode} and \ref{eq:wz}) and perturbation equations, is implemented numerically to enable confrontation with observational data. We incorporate this model into a modified version of the \textsf{CAMB} Boltzmann solver \citep{Lewis:1999bs}. Bayesian parameter estimation is subsequently performed using the \textsf{Cobaya}\footnote{\url{https://github.com/CobayaSampler/cobaya}} framework \citep{2021JCAP...05..057T}. The Markov Chain Monte Carlo (MCMC) sampling convergence is rigorously assessed via the Gelman-Rubin criterion \citep{1992StaSc...7..457G}, requiring $R-1 < 0.01$ for all parameters. We adopt uniform priors on the cosmological parameter set $\{\Omega_b h^2, \Omega_c h^2, \tau_{\text{reio}}, n_s, \log(10^{10}A_s), \Delta, z_t\}$, with the resulting posterior distributions analyzed using the \textsf{GetDist} package\footnote{\url{https://github.com/cmbant/getdist}}. This comprehensive approach enables the robust determination of one-dimensional marginalized constraints and two-dimensional joint confidence regions for all model parameters.

\begin{table*}[htbp]
\label{tab:cosmo_params1}
  \centering
  \caption{Constraints on cosmological parameters for the flat MEDE model. Results are shown for the CMB+DESI, CMB+DESI+DES-Dovekie, and CMB+DESI+PantheonPlus datasets, and we also include the minimum $\chi^2$, $\Delta\chi^2$, and $\Delta \rm{DIC}$ values relative to the $\Lambda$CDM and CPL models.
  }

  \begin{tabular}{c|ccc}
    \toprule
    Model & \multicolumn{3}{|c}{MEDE model} \\
    \midrule
    Data & CMB+DESI & CMB+DESI+DES-Dovekie & CMB+DESI+PantheonPlus \\
    \midrule
$z_\mathrm{t}$          &  $0.500^{+0.130}_{-0.110}$ & $0.418^{+0.070}_{-0.085}$ & $0.425^{+0.084}_{-0.120}$\\
$\Delta$                &  $> 1.12$ & $1.03\pm 0.32$ & $0.87^{+0.29}_{-0.35}$\\
$\Omega_\mathrm{b} h^2$ &  $0.02226\pm 0.00012$ & $0.02226\pm 0.00012$ & $0.02226\pm 0.00013$\\
$\Omega_\mathrm{c} h^2$ &  $0.11895\pm 0.00078$ & $0.11902\pm 0.00077$ & $0.11890\pm 0.00081$\\
$\log(10^{10} A_\mathrm{s})$  $3.046\pm 0.012$ & $3.045\pm 0.012$ & $3.046\pm 0.012$\\
$n_\mathrm{s}$          &  $0.9662\pm 0.0036$ & $0.9661\pm 0.0035$ & $0.9664\pm 0.0035$\\
$\tau_\mathrm{reio}$    & $0.0564\pm 0.0067$ & $0.0561\pm 0.0062$ & $ 0.0567\pm 0.0066$\\
    \midrule
$H_0$  [km s$^{-1}$ Mpc$^{-1}$] &  $66.30^{+1.20}_{-1.40}$ & $67.43\pm 0.56$ & $67.56\pm 0.58$\\
$\Omega_\mathrm{m}$             &  $0.3230\pm 0.0120$ & $0.3122\pm 0.0055$ & $0.3108\pm 0.0056$\\
$\sigma_8$        & $0.8020\pm 0.0120$ & $0.8116\pm 0.0073$ & $0.8120\pm 0.0079$\\
$r_\mathrm{drag}$ &  $147.51\pm 0.21$ & $147.49\pm 0.20$ & $147.51\pm 0.20$\\
    \midrule
$\chi^2_{\mathrm{min}}$ &  $11006.35$ & $12632.67$ & $12406.45$\\
$\Delta \chi^2_{\text{MEDE-$\Lambda$CDM}}$ &  $-1.65$ & $-8.43$ & $-3.94$\\
$\Delta \chi^2_{\text{MEDE-CPL}}$&  $4.22$ & $4.53$ & $3.39$\\
    \midrule
$\Delta \rm{DIC}_{\text{MEDE-$\Lambda$CDM}}$ &  $-6.53$ & $-11.36$ & $-9.29$\\
$\Delta \rm{DIC}_{\text{MEDE-CPL}}$ &  $1.71$ & $1.32$ & $0.74$\\
    \bottomrule
    \end{tabular}
\end{table*}

\begin{figure}
\centering
\includegraphics[width=0.455\textwidth]{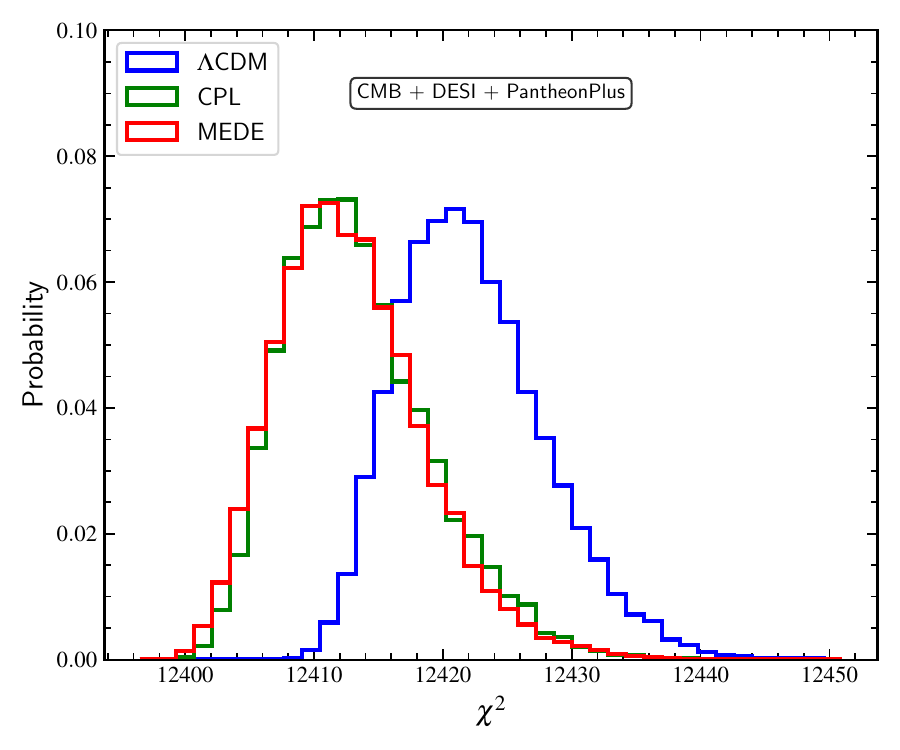}
\caption{ $\chi^2$ distributions from MCMC chains for $\Lambda$CDM (blue), 
        CPL dynamical dark energy (green), and MEDE (red) models, 
        constrained by CMB+DESI DR2+PantheonPlus data.}
\label{Fig:chi2}
\end{figure}

 {We employ the Deviance Information Criterion (DIC) \citep{10.1111/1467-9868.00353,10.1111/j.1745-3933.2007.00306.x} for model comparison. DIC is defined as
\begin{equation}
    \mathrm{DIC} = \bar{D} + p_D,
\end{equation}
where \(\bar{D}\) is the posterior mean of the deviance \(D(\theta) = -2\ln\mathcal{L}(\theta)\), and \(p_D = \bar{D} - D(\bar{\theta})\) is the effective number of parameters. By using the effective number of parameters, the DIC overcomes the problem of other information criteria like Akaike criterion, that it does not discount parameters which are unconstrained by the data. Under the Gaussian likelihood assumption, \(D(\theta)\) is equivalent to \(\chi^2(\theta)\). Thus, DIC can be expressed in terms of \(\chi^2\) as
\begin{equation}
\mathrm{DIC} = 2\langle \chi^2(\theta) \rangle - \chi^2(\bar{\theta}),
\end{equation}
where \(\langle \chi^2(\theta) \rangle\) is the mean of \(\chi^2\) over the posterior samples, and \(\chi^2(\bar{\theta})\) is \(\chi^2\) evaluated at the posterior mean of the parameters. A smaller DIC indicates a better balance between goodness-of-fit and model complexity.}

Our analysis of the MEDE model, combining the latest CMB, DESI BAO, and SN Ia (PantheonPlus and DES-Dovekie) datasets, yields robust constraints on key cosmological parameters. The 1D marginalized posterior distributions and 2D joint confidence regions are displayed in Fig.~\ref{Fig:mede1}, while numerical results are summarized in Table 1. For comparative assessment, Figure~\ref{Fig:chi2} displays the posterior distributions of the total $\chi^2$ 
for the three cosmological models under consideration, obtained from the joint 
CMB+DESI DR2+PantheonPlus analysis.

The most striking evidence for the emergent nature of dark energy comes from the highly consistent constraint on the transition redshift $z_t$ obtained from two complementary dataset combinations: $z_t = 0.418^{+0.070}_{-0.085}$ (CMB+DESI+DES-Dovekie) and $z_t = 0.425^{+0.084}_{-0.120}$ (CMB+DESI+PantheonPlus). This remarkable agreement strongly indicates that the dark energy transition occurs squarely within the epoch of cosmic acceleration onset ($z \sim 0.5$), suggesting a fundamental connection between the emergence of dark energy dominance and the recent acceleration phase. Furthermore, the positive values of $\Delta=1.03 \pm 0.32$ and $0.87^{+0.29}_{-0.35}$ from the respective datasets---consistently indicate a specific evolutionary pattern: dark energy evolves from phantom-like behavior ($w < -1$) at $z > z_t$ to quintessence-like behavior ($w > -1$) at $z < z_t$. This crossing of the phantom divide, while challenging for simple scalar field models, arises naturally in the MEDE framework through its hyperbolic parametrization, which can be mapped onto quintom scenarios with non-canonical kinetic terms. The concordance between these independent datasets significantly reinforces the robustness of this dynamical dark energy signature.

The addition of DESI BAO data serves as a powerful low-redshift anchor, breaking geometric degeneracies through its precise measurements of $D_M(z)/r_d$ and $H(z)r_d$ across multiple redshifts. Within the CMB+DESI dataset, this leads to tightly constrained parameters, with $H_0 = 66.30^{+1.20}_{-1.40}$ km s$^{-1}$ Mpc$^{-1}$ and a transition parameter $\Delta > 1.12$. The BAO distance measurements strongly constrain the sound horizon scale $r_{\rm drag} = 147.51 \pm 0.21$ Mpc, which remains highly consistent across different data combinations - further attesting to the stability of early-universe physics in the MEDE framework. The MEDE model also shows improved fit quality over $\Lambda$CDM in this dataset, as reflected by $\Delta\chi^2_{\text{MEDE-$\Lambda$CDM}} = -1.65$, $\Delta\mathrm{DIC}_{\text{MEDE-$\Lambda$CDM}} = -6.53$.

The subsequent inclusion of DES-Dovekie and PantheonPlus supernova data further refines the measurement of the cosmic expansion history. Supernovae provide direct observations of the luminosity distance, which integrates over the Hubble function $H(z)$. Remarkably, the parameter constraints from the CMB+DESI+DES-Dovekie and CMB+DESI+PantheonPlus combinations show a high degree of consistency: the transition parameter $\Delta = 1.03 \pm 0.32$ and $0.87^{+0.29}_{-0.35}$, the Hubble constant $H_0 = 67.43 \pm 0.56$ km s$^{-1}$ Mpc$^{-1}$ and $67.56 \pm 0.58$ km s$^{-1}$ Mpc$^{-1}$, and the matter density parameter $\Omega_{\mathrm{m}} = 0.3122 \pm 0.0055$ and $0.3108 \pm 0.0056$, respectively. This consistency underscores the robustness of the MEDE framework against different cosmological probes, especially given the distinct systematics and redshift coverages of the DES-Dovekie and PantheonPlus samples. Furthermore, the addition of supernova data leads to a significant improvement in the goodness-of-fit for the MEDE model over $\Lambda$CDM: for the CMB+DESI+DES-Dovekie combination, $\Delta\chi^{2}_{\text{MEDE-$\Lambda$CDM}} = -8.43$ and $\Delta\mathrm{DIC}_{\text{MEDE-$\Lambda$CDM}} = -11.36$; for the CMB+DESI+PantheonPlus combination, $\Delta\chi^{2}_{\text{MEDE-$\Lambda$CDM}} = -3.94$ and $\Delta\mathrm{DIC}_{\text{MEDE-$\Lambda$CDM}} = -9.29$. These results collectively indicate a preference for the MEDE model over $\Lambda$CDM when late-time observational data are incorporated.

Compared with the widely adopted CPL parametrization, the statistical evidence remains inconclusive. The positive $\Delta\chi^{2}_{\text{MEDE-CPL}}$ values indicate that, in terms of the minimum $\chi^2$, the CPL parametrization provides a slightly better fit to the data than the MEDE model. This is further supported by the DIC: since a lower DIC value is preferred, the positive $\Delta\mathrm{DIC}_{\text{MEDE-CPL}}$ values imply that, after accounting for model complexity, the data show a mild preference for CPL over MEDE. It is important to note, however, that all $\Delta\mathrm{DIC}$ values fall below the conventional threshold of 5 for substantial evidence. Therefore, current data do not provide strong statistical grounds to decisively distinguish between the MEDE and CPL models.

The stability of primordial parameters across different dataset combinations demonstrates that the MEDE model retains the success of $\Lambda$CDM in describing early-universe physics, while introducing additional dynamical freedom only at late times. This clear separation of scales is a desirable feature for any dark-energy extension, as it avoids interference with the well-constrained physics of the primordial universe.

Overall, the MEDE framework presents a compelling, theoretically motivated alternative to $\Lambda$CDM, as evidenced by the negative $\Delta\chi^{2}$ and $\Delta\mathrm{DIC}$ values relative to the standard model. Its performance is comparable to the widely used CPL parametrization, with the small $\Delta\mathrm{DIC}$ values indicating no strong statistical preference for either model. This positions MEDE as a viable dynamical dark-energy scenario worthy of further investigation with future, more precise observational data.

\section{Results and Discussion}\label{sec:results}
\begin{figure*}
\centering
\includegraphics[width=0.455\textwidth]{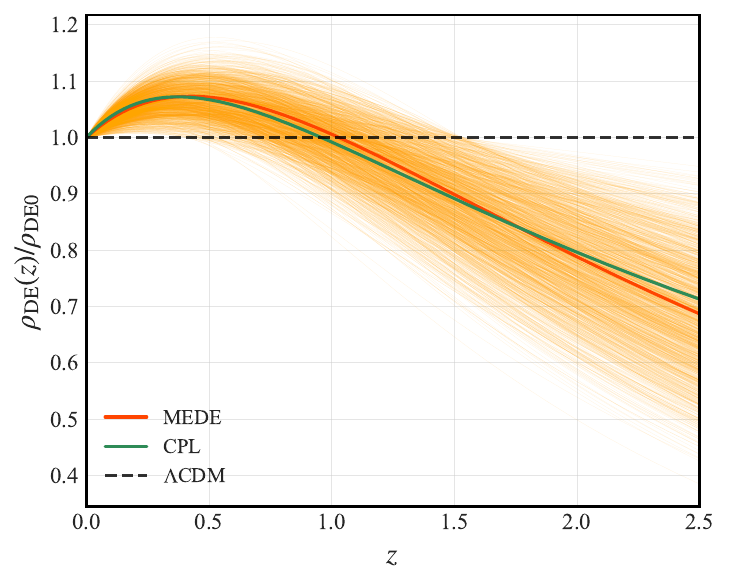}
\includegraphics[width=0.455\textwidth]{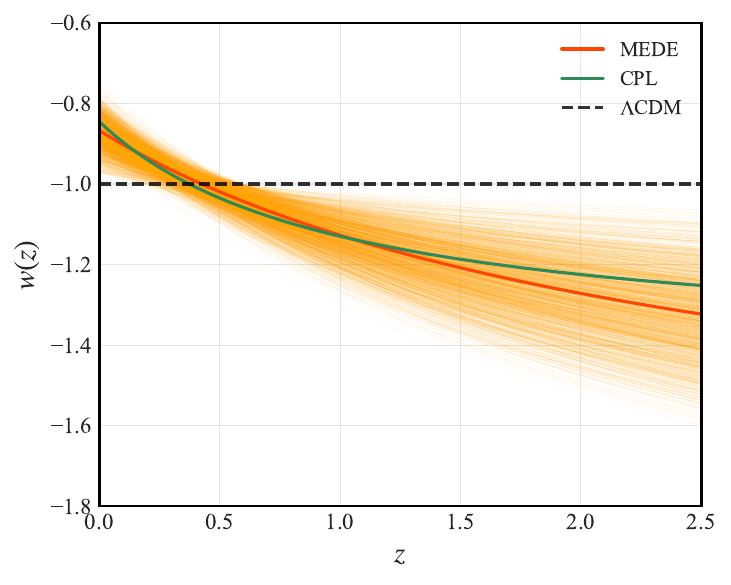}
\caption{Redshift evolution of the dark energy sector.
 {Left:} The dark energy density $\rho_{\mathrm{DE}}(z)/\rho_{DE0}$.
 {Right:} The equation of state $w_{\mathrm{DE}}(z)$.
The red solid line represents the best-fit MEDE model (CMB+DESI+PantheonPlus). The light orange region shows the $2\sigma$ confidence region derived from MCMC chains, illustrating the posterior distribution of parameters. The green solid line shows the CPL parametrization ((CMB+DESI+PantheonPlus best-fit), and the black dashed line represents $\Lambda$CDM. The MEDE model exhibits a smooth transition around $z_t \sim 0.42$, crossing the phantom divide $w = -1$ from phantom-like ($w < -1$) at high redshifts to quintessence-like ($w > -1$) at low redshifts.}
\label{Fig:mede3}
\end{figure*}

\begin{figure*}
\centering
\includegraphics[width=0.455\textwidth]{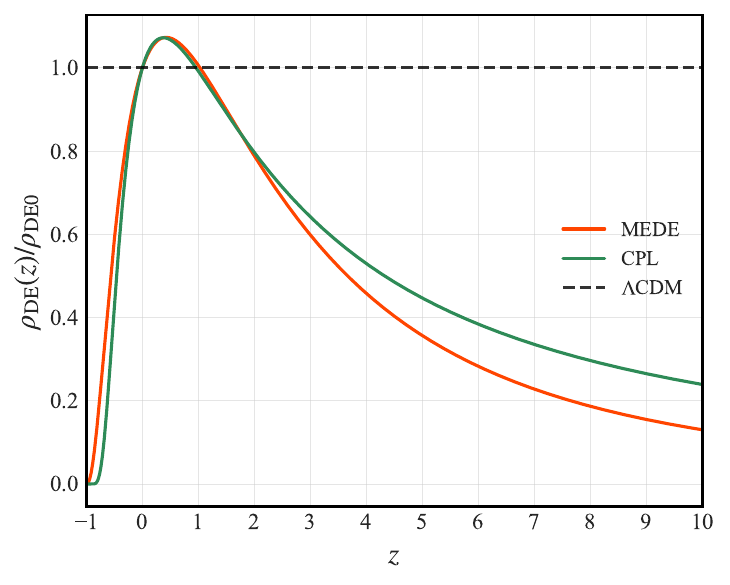}
\includegraphics[width=0.455\textwidth]{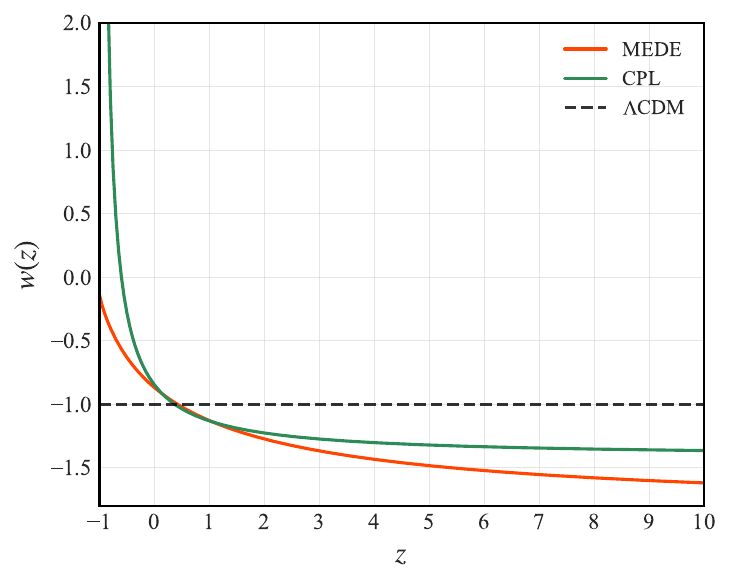}
\caption{ {Redshift evolution of the dark energy sector for MEDE, CPL, and $\Lambda$CDM.
Left: Dark energy density $\rho_{\mathrm{DE}}(z)/\rho_{\mathrm{DE},0}$.
Right: Equation of state $w_{\mathrm{DE}}(z)$.
The red solid line shows the best-fit MEDE model (CMB+DESI+PantheonPlus).
The green solid line is the best-fit CPL parametrization, and the black dashed line is $\Lambda$CDM. The evolution is shown from the distant past ($z \approx 10$) to the asymptotic future ($z \to -1$), illustrating the distinct behaviors of each model at late times.}}
\label{Fig:mede-bf}
\end{figure*}

Figure~\ref{Fig:mede3} illustrates the redshift evolution of the dark energy sector in the MEDE framework. The left panel shows the dark energy density $\rho_{\mathrm{DE}}(z)/\rho_0$, while the right panel displays the corresponding equation of state $w_{\mathrm{DE}}(z)$. The red solid line represents the best-fit MEDE model from the CMB+DESI+PantheonPlus dataset, with $z_t = 0.425$ and $\Delta = 0.87$, while the surrounding light orange region indicates the $2\sigma$ confidence levels derived from MCMC chains. For comparison, the CPL parametrization with $w_0 = -0.845$ and $w_a = -0.57$ best fit from  CMB+DESI+PantheonPlus dataset is shown as a green solid line, and the $\Lambda$CDM limit is shown as a black dashed line. The constrained transition around $z_t \sim 0.42$ reveals a smooth crossing of the phantom divide ($w = -1$), where dark energy evolves from phantom-like behavior ($w < -1$) at high redshifts to quintessence-like behavior ($w > -1$) at low redshifts. This figure visually demonstrates the unique evolutionary pattern of the MEDE model, which naturally accommodates phantom crossing while maintaining theoretical consistency across the entire redshift range.

 { Figure~\ref{Fig:mede-bf} shows the redshift evolution of the dark energy sector for MEDE, CPL, and $\Lambda$CDM, with the left panel displaying the dark energy density $\rho_{\mathrm{DE}}(z)/\rho_{\mathrm{DE},0}$ and the right panel the equation of state $w_{\mathrm{DE}}(z)$. As illustrated, the MEDE equation of state (best fit to the current data) remains well behaved asymptotically into the infinite future, whereas CPL can evolve to values beyond $w=1$, which are difficult to interpret physically. This highlights an important difference: MEDE is constructed from a physically motivated density evolution, in which dark energy appears and then disappears, rather than from an assumed form of the equation of state.}

 {Beyond statistical comparison, it is also instructive to consider how these parametrizations relate to underlying physical models. Studies such as \cite{Wolf:2023uno,Bayat:2025xfr,Wolf:2024eph,Shlivko:2025fgv} show that CPL may struggle to reconstruct the true equation of state of certain scalar field models, while \cite{Wolf:2025jlc} demonstrates it can still reproduce key observables within current uncertainties. More generally, the limitations of fixed parametric forms have long motivated non-parametric approaches: for example, CPL fails to fit certain braneworld models \citep{Shafieloo:2005nd} and cannot adequately capture sharp transitions in the equation of state \citep{2010PhRvD..82j3502H}. CPL remains a flexible empirical tool, whereas MEDE is explicitly rooted in the emergent dark energy paradigm---preserving the emergence feature of PEDE/GEDE while allowing phantom crossing. Thus, MEDE should be seen not as a replacement for CPL, but as a complementary framework grounded in a specific physical picture.}

 {While both CPL and MEDE are currently phenomenological, CPL by construction has a more limited redshift range of validity than MEDE. Work is under way to ground MEDE in a concrete microphysical scenario involving scalar fields, which would further elevate it from a phenomenological proposal to a more fundamentally motivated one.}

\section{Conclusion} \label{sec:con}

We have proposed the Metastable Emergent Dark Energy (MEDE) model, a novel phenomenological extension of the Phenomenological (PEDE) and Generalized (GEDE) Emergent Dark Energy frameworks, in which dark energy exhibits both emergent and transitionary behavior, appearing dynamically at late times and potentially vanishing toward the future. This construction provides a simple yet flexible framework in which dark energy is not a rigid cosmological constant, but rather a metastable component whose effective contribution to the cosmic expansion can evolve non-trivially across cosmic time. In particular, the model naturally accommodates scenarios in which dark energy effectively fades away in the far future, thereby opening qualitatively new perspectives on the ultimate fate of the Universe.

The MEDE model is defined by the hyperbolic equation of state
$w(z)=-1-\Delta\tanh[\log_{10}((1+z)/(1+z_t))]$,
which introduces only two additional free parameters beyond $\Lambda$CDM, namely the transition redshift $z_t$ and the variation amplitude $\Delta$. Despite its minimal parametrization, the model successfully captures both the emergent onset of dark energy domination and a transitionary behavior that allows for a smooth and physically motivated crossing of the phantom divide line. Such phantom-crossing behavior has been hinted at by several recent low-redshift observations from BAO and supernovae, and MEDE provides a coherent phenomenological realization of this possibility without introducing pathologies at the background level.

Constraining the model using a combined dataset of Planck CMB, DESI DR2 BAO, and different compilations of Type Ia supernovae, we obtain $z_t=0.425^{+0.084}_{-0.120}$ and $\Delta =0.87^{+0.29}_{-0.35}$ (for CMB+DESI+PantheonPlus), indicating a statistically significant deviation from the cosmological constant scenario. Statistical comparisons show that the MEDE model is preferred over $\Lambda$CDM by the combined dataset, with $\Delta \rm DIC_{ MEDE-\Lambda CDM}= -9.29$, while it performs comparably to the CPL dynamical dark energy parametrization ($\Delta \rm DIC_{MEDE-CPL} = 0.74$), with no strong statistical distinction from CPL using current data. In this sense, MEDE emerges as a competitive and equally viable alternative to CPL and other compelling dynamical dark energy models.

Importantly, the MEDE framework preserves the well established successes of $\Lambda$CDM in describing early-universe physics, while extending the late-time sector in a controlled and testable manner. The combination of emergent behavior, transitionary dynamics, and phantom-crossing capability makes MEDE a particularly appealing phenomenological scenario. Beyond its empirical performance, the model may serve as guidance toward more fundamental theoretical constructions, potentially pointing to underlying metastable vacuum structures, interacting sectors, multi-field models, or modified gravitational dynamics responsible for the effective behavior captured by the parametrization.  {Owing to space limitations, the physical origin framework, action, and consistency check are not covered in this paper. A comprehensive presentation of these aspects of the MEDE model will be provided in future work.}

Looking ahead, forthcoming high precision cosmological observations will be crucial in clarifying the current hints of dynamical dark energy. Future surveys with improved sensitivity to low-redshift expansion history and structure growth will be able to discriminate more decisively between $\Lambda$CDM, CPL, MEDE, and other competing models. In this context, the MEDE scenario provides a minimal, flexible, and observationally testable framework that stands as a competitive alternative in the exploration of the nature and fate of dark energy.

\section*{Acknowledgements}
This work was supported  by  the National Key R\&D Program of China (No. 2024YFC2207400) and the Science Research Project of Hebei Education Department No. BJK2024134.
This work benefits from the high performance computing clusters at College of Physics, Hebei Normal University; The Chutian Scholars Program in Hubei Province (X2023007); A. S. would like to acknowledge the support by National Research Foundation of Korea NRF2021M3F7A1082056 and the support of the Korea Institute for Advanced Study. M.B. was supported by the Polish National Science Centre grant 2023/50/A/ST9/00579 and by COST Action CA21136 – “Addressing observational tensions in cosmology with systematics and fundamental physics (CosmoVerse)”.

\bibliography{main}{}
\bibliographystyle{aasjournal}
\end{document}